\documentclass[aps,prb,preprint,superscriptaddress,showpacs,showkeys]{revtex4}

\usepackage{graphicx}

\begin{document}

\title{Anomalous metallic state in quasi-two-dimensional BaNiS$_{2}$}

\author{David Santos-Cottin}
\email[]{david.santos-cottin@espci.fr}
\author{Andrea Gauzzi}
\author{Marine Verseils}
\author{Benoit Baptiste}
\affiliation{IMPMC, Sorbonne Universit\'e-UPMC, CNRS, IRD, MNHN, 4 place Jussieu, 75252 Paris, France}

\author{Gwendal Feve}
\author{Vincent Freulon}
\author{Bernard Pla\c cais}
\affiliation{Laboratoire Pierre Aigrain, Ecole Normale Sup\'erieure-PSL Research University, CNRS, Universit\'e Pierre et Marie Curie-Sorbonne Universit\'es, Universit\'e Paris Diderot-Sorbonne Paris Cit\'e, 24 rue Lhomond, 75231 Paris Cedex 05, France}

\author{Michele Casula}
\author{Yannick Klein}
\email[]{yannick.klein@upmc.fr}
\affiliation{IMPMC, Sorbonne Universit\'e-UPMC, CNRS, IRD, MNHN, 4 place Jussieu, 75252 Paris, France}

\date{\today}

\begin{abstract}
We report on a systematic study of the thermodynamic, electronic and charge transport properties of high-quality single crystals of BaNiS$_2$, the metallic end-member of the quasi-twodimensional BaCo$_{1-x}$Ni$_x$S$_2$ system characterized by a metal-insulator transition at $x_{cr}=0.22$. Our analysis of magnetoresistivity and specific heat data consistently suggests a picture of compensated semimetal with two hole- and one electron-bands, where electron-phonon scattering dominates charge transport and the minority holes exhibit, below $\sim$100 K, a very large mobility, $\mu_h\sim$ 15000 cm$^2$V$^{-1}$s$^{-1}$, which is explained by a Dirac-like band. Evidence of unconventional metallic properties is given by an intriguing crossover of the resistivity from a Bloch-Gr\"uneisen regime to a linear$-T$ regime occurring at 2 K and by a strong linear term in the paramagnetic susceptibility above 100 K. We discuss the possibility that these anomalies reflect a departure from conventional Fermi-liquid properties in presence of short-range AF fluctuations and of a large Hund coupling.

\end{abstract}



\maketitle

\section{Introduction}

\begin{figure}[!h]
   \includegraphics[width=10 cm]{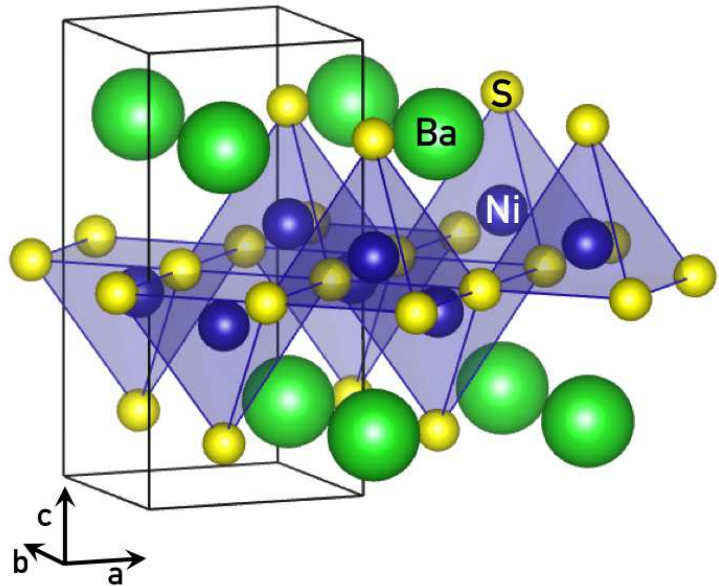}
   \caption{\label{structure} (color online) Tetragonal structure of BaNiS$_2$. The unit cell is drawn using solid lines.}
\end{figure}

BaNiS$_2$ is the metal phase precursor of the metal-insulator transition (MIT) observed at $x_{cr}=0.22$ in the quasi-two-dimensional square-lattice BaCo$_{1-x}$Ni$_x$S$_2$ system, where the Ni/Co substitution level, $x$, controls electron doping. This MIT has attracted interest for it is associated with a competition between an insulating antiferromagnetic (AF) phase and a paramagnetic metallic one \cite{tak94}, similar to the case of unconventional superconductors, such as cuprate \cite{ima98}, Fe-based \cite{joh10} and heavy-fermions \cite{joy02}. On the other hand, no superconductivity has been hitherto reported in BaCo$_{1-x}$Ni$_x$S$_2$. In its simple tetragonal $P4/nmm$ structure \cite{kod95} made of a checkerboard-like network of edge-sharing NiS$_5$ pyramids (see Fig.\ref{structure}), all atomic coordinates are set by symmetry in the $ab$-plane and the structural degrees of freedom are limited to the $z$-coordinates along the $c$-axis. Because of the absence of structural distortions concomitant to the MIT, BaCo$_{1-x}$Ni$_{x}$S$_{2}$ is a model system for studying the doping-controlled Mott transition in a square lattice \cite{lee06}, where the electronic degrees of freedom are decoupled from those of the lattice.

In order to unveil the mechanism of the MIT, previous studies have been mostly devoted to the doping region in the vicinity of $x_{cr}$, whilst the precursor metallic phase BaNiS$_{2}$ remains little studied. It is the purpose of the present work to systematically investigate this phase, which should help elucidating the stability of the metallic state in the important limit of no chemical disorder ($x$=1). Open questions are the relevance of electron-electron correlations to charge localization and the possibility of unconventional topological phases that may arise from the following features of the electronic structure of BaNiS$_2$: (i) strong spin-orbit (SO) coupling effects leading to a large Rashba splitting, which is unexpected for a compound composed by comparatively light element \cite{ARPES}; (ii) a Dirac-like point at the Fermi surface. In the present work, we carried out a systematic study by means of specific heat, susceptibility and magnetoresistivity measurements on high-quality single crystals. The data give evidence of anomalous properties suggesting a quantum critical point scenario controlled by AF fluctuations and by a strong Hund coupling involving the Ni$^{2+}$ ions.  
 
\section{Methods}

BaNiS$_{2}$ single crystals were synthesized using a conventional solid state reaction method, as described in detail elsewhere \cite{sha95}. In brief, powders of barium sulfide (BaS, 99.9\% purity), sulfur (99.995\%) and metallic nickel (99.999\%) in non-stoichiometric molar ratios Ba:Ni:S = 0.10:0.425:0.475 were finely ground and pelletized. The pellets were loaded in a graphite crucible and sealed in a quartz ampoule in vacuum at pressures of 10$^{-5}$ mbar or better. Graphite was used as a catalytic matrix for oxygen, thus preventing oxidation of the reagents. The heat treatment consists of a first dwell at 300 $^{\circ}$C for 2h followed by a second one at 1100 $^{\circ}$C for 48h, a cooling down to 850 $^{\circ}$C at a rate of 50 $^{\circ}$C/h and a final quench into water. The powders were ground again and a similar heat treatment as before was applied, except the cooling rate was reduced to 1 $^{\circ}$C/h. This second treatment leads to the formation of black platelet-like single crystals of size up to $\sim 1 \times 1 \times 0.1$ mm$^3$ that were mechanically removed from the batch and washed with ethanol. Several single crystals were selected for the present study. Single crystal x-ray diffraction yielded structural parameters \textit{a} = \textit{b} = 4.4404(6) \AA, and \textit{c} = 8.897(2) \AA, in agreement with previous results \cite{kod95,gre70}. Structural refinements carried out assuming no sulfur or nickel vacancies yielded exceptionally low reliability factors, $R_w < 0.015$, thus indicating a low amount of disorder. Refinements that include the possibility of vacancies did not improve the result and systematically converged to the stoichiometric formula. 

\indent
Magnetization and specific heat measurements were carried out in a Quantum Design SQUID vibrating sample magnetometer (VSM) at a field of 1 T and in a Quantum Design physical properties measurement system (PPMS) using a 2-$\tau$ relaxation method, respectively. For transport measurements, the samples were contacted with silver epoxy (Dupont 6838 conductor paste) which ensures low-resistance ohmic contacts after a heat treatment under vacuum at 250 $^{\circ}$C overnight. The in-plane ($ab$) longitudinal and transverse resistances, \textit{R}$_{xx}$ and \textit{R}$_{xy}$ respectively, were measured using a AC four-probe technique in the PPMS in the bar or van der Pauw \cite{VDP} configurations. The out-of-plane (\textit{c}-axis) resistivity was measured by employing a ring geometry of the current electrode, as described elsewhere \cite{c-axis}. The in-plane resistivity of a representative single crystal was measured at very low temperatures down to 40 mK in a He$^3$ dilution refrigerator cryostat using a lock-in detection technique.

\indent
The band structure has been calculated with the Perdew-Burke-Ernzerhof density functional \cite{PBE} augmented by a local Coulomb repulsion via the spherically symmetric Hubbard matrix with $U =$ 3 eV and by a Hund coupling $J =$ 0.95 eV, while the Racah parameter ratio $F_4/F_2$ is taken from the atomic value. We used norm-conserving pseudopotentials by treating the Ni pseudoatom as fully relativistic. The spin-orbit coupling is taken into account by using the Quantum Espresso implementation \cite{espresso}, with a spinor formulation based on non-collinear two-component spin-Bloch functions \cite{Baroni,DalCorso}. The self-consistent calculation has converged on a 8x8x8 Monkhorst-Pack $k-$points grid with a plane-waves cutoff of 120 Ry. The density of states has been computed with a non-self consistent calculation on a 16x16x16 $k-$points grid and the tetrahedron method \cite{tetrahedron}.  

\section{Results}

\begin{figure}[!h]
\includegraphics[width=10 cm]{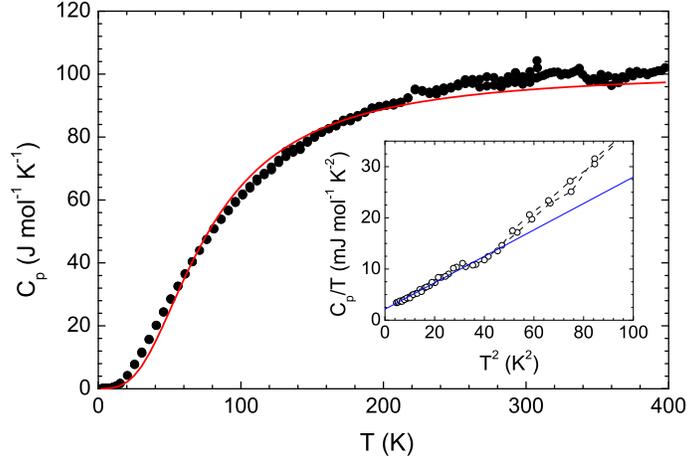}
\caption{\label{Cp} (color online) Temperature dependence of the isobaric specific heat, $C_P$, of a bunch of BaNiS$_2$ single crystals. In the inset, the broken line is a fit of the low temperature data that yields $\gamma$ and $\Theta_D$. The red solid line is the theoretical Debye curve described by Eq. \ref{Debye}.}    
\end{figure}

Fig. \ref{Cp} shows the temperature-dependent isobaric specific heat, $C_P(T)$, measured on a bunch of single crystals of total weight 5.77 mg selected from the same batch. No anomalies are found in the whole 2-400 K range measured, which indicates the absence of phase transitions. As customary done for solids in the above range, in the present data analysis, we neglect the small difference between the \textit{isobaric} and \textit{isochoric} specific heats given by $C_P-C_V=\alpha^2K_T V T$, where $\alpha$, $K_T$ and $V$ are the thermal expansion coefficient, the isothermal bulk modulus and the molar volume, respectively. Considering the experimental value of $\alpha$ reported for BaNiS$_2$ \cite{tak97} and typical values for $K_T$, the magnitude of this difference in the temperature range considered here is estimated to be $\sim$1-2 \% or less. We then analyze the data using the usual expression for $C_V$ that reads

\begin{equation}
\label{Debye}
C_V(T)=\gamma T+9sk_B\left(\frac{T}{\theta_D}\right)^3\int_0^{\theta_D/T}
\frac{x^4 e^{x}}{(e^{x} -1)^2}dx
\end{equation}
\\
where the first and second terms represent the electronic and lattice contributions, respectively, described by a conventional band model and by the Debye model. The quantities $\gamma$, $s$, $k_B$ and $\theta_D$ denote the Sommerfeld coefficient, the number of atoms per mole, the Boltzmann constant and the Debye temperature, respectively. By fitting the low temperature data below 8 K, where the expression in Eq. \ref{Debye} is approximated by $C_V(T)/T = \gamma + \beta T^2$ and $\beta \propto \theta_D^{-3}$, we obtain $\gamma=2.15(11)$ mJ K$^{-2}$ mol$^{-1}$ and $\theta_D = 311(5)$ K. This $\gamma$ value is slightly larger than previously measured on polycrystalline samples \cite{tak95} but comparable to the values 2.9 mJ K$^{-2}$ mol$^{-1}$ \cite{mat95} and 3.1 mJ K$^{-2}$ mol$^{-1}$ \cite{has95} estimated from \textit{ab initio} DFT calculations of the density of states (DOS) at the Fermi level, $D(\epsilon_F)$, which suggests a modest mass renormalization. The extrapolation of the Debye curve to higher temperatures using the above parameters accounts for the data up to 400 K. The slight excess of the experimental value with respect to the Dulong-Petit limit $3sk_B=99.8$ J mol$^{-1}$ K$^{-1}$ observed at such higher temperatures, is ascribed to the small difference between $C_V$ and $C_P$.

\begin{figure}[!h]
\includegraphics[width=10 cm]{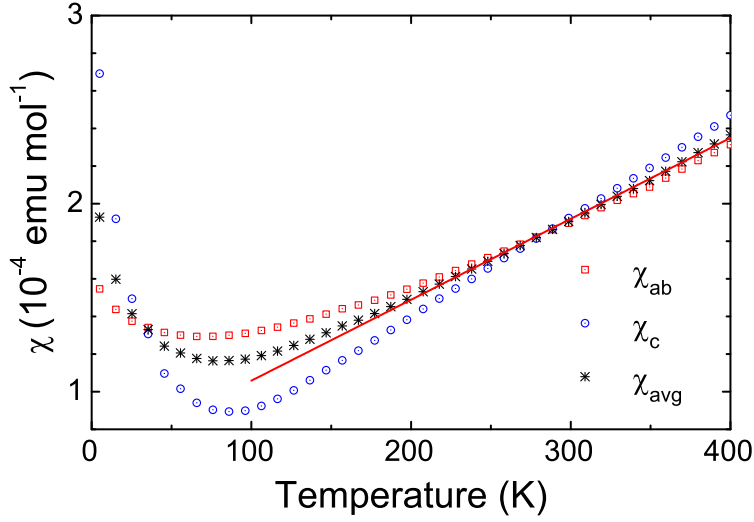}
\caption{\label{susceptibility} (color online) Magnetic susceptibility of a BaNiS$_2$ single crystal. $\chi_{ab}$ and $\chi_c$ indicate the measurements taken with field in the \textit{ab}-plane and along the \textit{c}-axis, respectively. The average is given by $\chi_{avg} = \frac{2}{3}\chi_{ab} + \frac{1}{3}\chi_c$. The data have been corrected from the core diamagnetic susceptibility estimated to be $\chi_{core} \approx -1.2 \times 10^{-4}$ emu mol$^{-1}$.}   
\end{figure}

In Fig. \ref{susceptibility}, we plot the DC magnetic susceptibility curves, $\chi_{ab}(T)$ and $\chi_{c}(T)$, in fields parallel and perpendicular to the \textit{ab}-plane for a representative BaNiS$_2$ single crystal. The difference between the two curves indicates a pronounced anisotropy of the magnetic response. The averaged susceptibility is in good agreement with previous results on polycrystalline samples \cite{kur93,tak94,mar95}, except the present data show a weaker upturn at low temperature. At first sight, this may be due to a smaller concentration of paramagnetic impurities in the present single crystals. Though, no clear evidence of Curie-like behavior is found at low temperatures, so the physical origin of the upturn may be different. Notable feature of both $\chi_{ab}(T)$ and $\chi_{c}(T)$ is a pronounced linear behavior at high temperature with a large slope $4.2 \times 10^{-7}$ emu mol$^{-1}$ K$^{-1}$. Such behavior contrasts the expectation of a constant Pauli term in a conventional metal. A similar linear dependence with comparable slopes $\sim 4.3 \times 10^{-7}$ emu K$^{-1}$ mol(Fe)$^{-1}$ and $\sim 6 \times 10^{-7}$ emu K$^{-1}$ mol(Fe)$^{-1}$ has been previously reported in other semimetals, such as the parent compounds of Fe-based superconductors LaOFeAs and BaFe$_{2}$As$_{2}$, respectively \cite{kli10,wan09}. For these systems, two scenarios have been proposed, one based on AF fluctuations and the other on a pronounced peak of the density of states at $\epsilon_F$. Below, we shall discuss the suitability of these two scenarios for the present BaNiS$_2$ case.
 
\begin{figure}[!h]
   \includegraphics[width=10 cm]{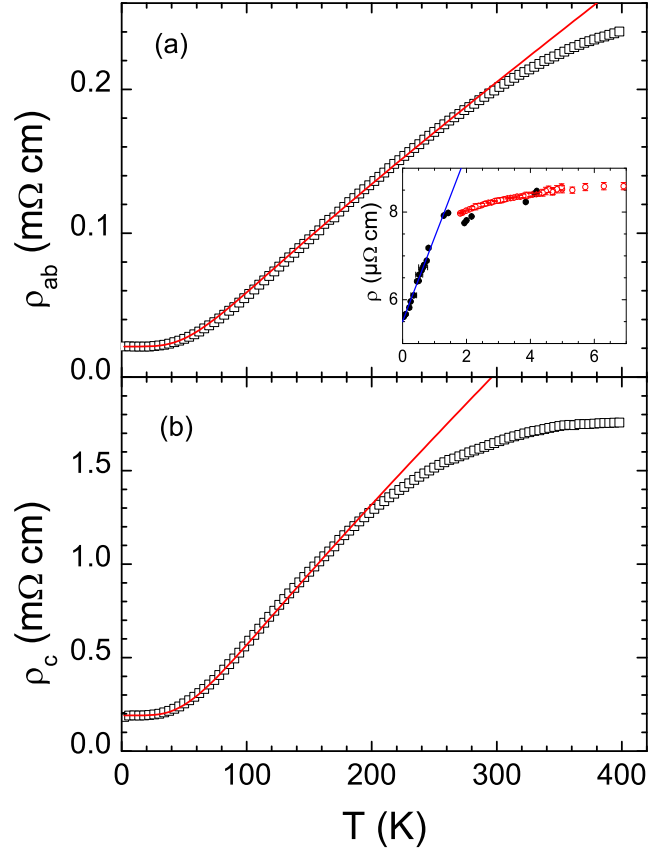}
   \caption{\label{resistivity} \textit{ab}-plane (a) and \textit{c}-axis (b) resistivity of a representative BaNiS$_2$ single crystal with $RRR = 9.4$. The red lines are a data fit using eq. (\ref{BG}). The inset shows the low temperature resistivity curve obtained on a sample with $RRR$=15.8. Black and red circles indicate data obtained in a dilution cryostat and in a PPMS, respectively. The blue line is a linear fit $\rho = \rho'_0+AT$ with $A = 1.91 \times 10^{-3}$ m$\Omega$cm K$^{-1}$ and $\rho'_0 = 5.5 \times 10^{-3}$ m$\Omega$cm.}
\end{figure}

Figure \ref{resistivity} shows the \textit{ab}-plane and \textit{c}-axis components of the electrical resistivity tensor, $\rho_{ab}(T)$ and $\rho_c(T)$. $\rho_{ab}(T)$ was reproducibly measured on a dozen single crystals with residual resistivity ratios $RRR=\rho_{ab}$(300K)/$\rho_0$ ranging from 4 up to 17, where $\rho_0$ denotes the residual resistivity measured at 10 K (see below and inset of Fig. \ref{resistivity}). The largest $RRR$'s are about 4 times larger than those previously reported \cite{sha95}, which indicates a higher purity of the present crystals. The results of the present measurements confirm previous work, in particular the value of the room temperature resistivity $\rho_{ab}$(300K) $\approx$ 0.2 m$\Omega$cm, characteristic of a bad metal. $\rho_{ab}$ and $\rho_{c}$ are successfully explained by a conventional Bloch-Gr\"{u}neisen model of electron-phonon scattering described by the expression: 

\begin{equation}
\label{BG}
\rho(T)=\rho_0+\alpha \left(\frac{T}{\theta_D}\right)^5\int_0^{\theta_D/T}
\frac{x^5 e^{x}}{(e^{x} -1)^2}dx
\end{equation}
\\
where the Eliashberg function $\alpha \sim \lambda \theta_D/\omega_p^2$, assumed to be temperature independent, depends upon the electron-phonon coupling constant, $\lambda$, and upon the plasma frequency, $\omega_p$ \cite{BG}. The above expression is found to account well for both $\rho_{ab}$ and $\rho_{c}$ curves up to 250 K and 200 K, respectively. A data fit yields almost identical values for the Debye temperature, $\theta_D = 330$ K for $\rho_{ab}$ and $\theta_D = 332$ K for $\rho_{c}$. These values are in very good agreement with the $\theta_D =311$ K obtained from the specific heat data, which strongly supports a picture of electron-phonon scattering. The absence of a $T^2$ term at low temperature indicates negligible electron-electron scattering, in agreement with the previous observation of small mass renormalization effects by means of specific heat (see above) and ARPES measurements \cite{ARPES}. The anisotropy ratio is found to be $\kappa=\rho_c/\rho_{ab}\sim$8 at 300 K and does not depends strongly on temperature and may vary in the 6-10 range from sample to sample. These values suggest quasi-twodimensional transport properties; a more pronounced 2D character is found in unconventional metals, such as Sr$_2$RuO$_4$ ($\kappa\approx 200$), BaFe$_2$As$_2$ ($\kappa\approx 150$) and underdoped cuprates, such as YBa$_{2}$Cu$_3$O$_{7-\delta}$ ($\kappa\approx 100$) \cite{mae94,wan09,Zve03}. Our observation of a significant contribution of the $c$-axis conductivity is consistent with the existence of dispersive bands along $k_z$ \cite{ARPES,has95,mat95}. The present result on single crystal contrasts a previous result on \textit{c}-axis oriented polycrystalline samples in which much larger ratios $\kappa\sim10^3$ were reported \cite{kur93}. This discrepancy may arise from the enhanced charge scattering at grain boundaries.

Notable are the following anomalies of the temperature dependence of the resistivity:

\begin{enumerate}
\item 
At high temperature, the resistivity tends to level off.

\item
In the highest purity samples, below $T^{\ast} \sim 6$ K, the in-plane resistivity does not level off at the value of the residual resistivity, $\rho_0$, as in a normal metal; instead, it begins decreasing linearly with decreasing temperature down to the lowest temperature measured, 40 mK (see inset of Fig. \ref{resistivity}a).
\end{enumerate}

The possible origin of these two anomalies are discussed below. 

\begin{figure}[!h]
   \includegraphics[width=10 cm]{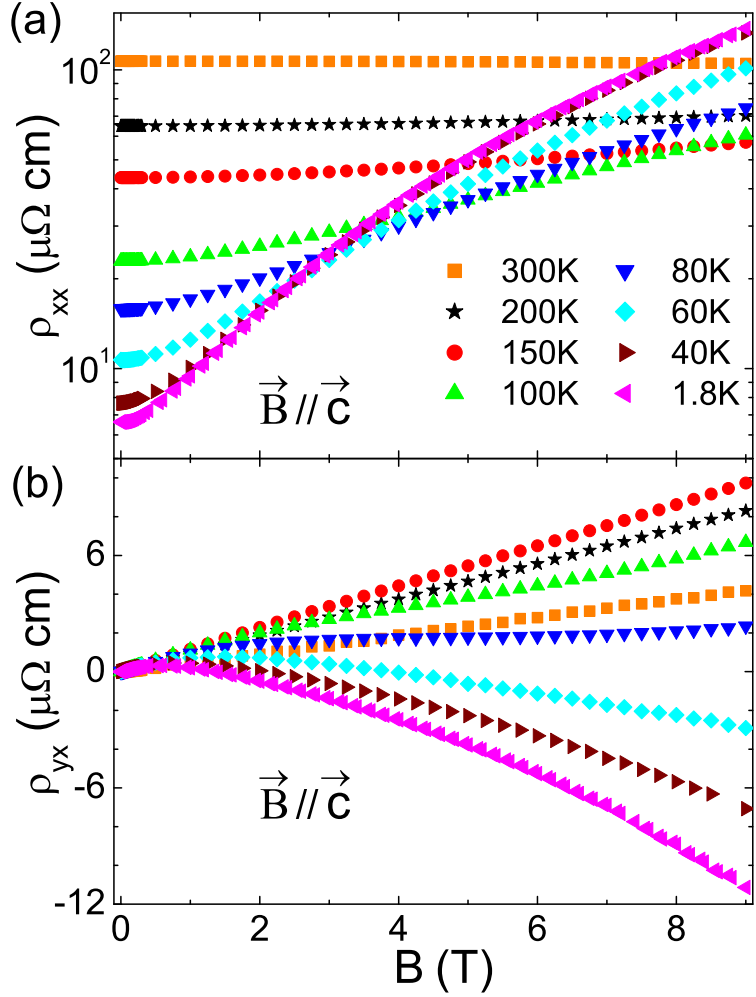}
   \caption{\label{Mresistivity} (a) Transverse and (b) longitudinal magnetoresistivity of a BaNiS$_2$ single crystal with RRR = 15.8 at temperatures ranging from 1.8 K to 300 K. The magnetic field was applied perpendicular to the current.}
\end{figure}

\begin{figure}
   \includegraphics[width=10 cm]{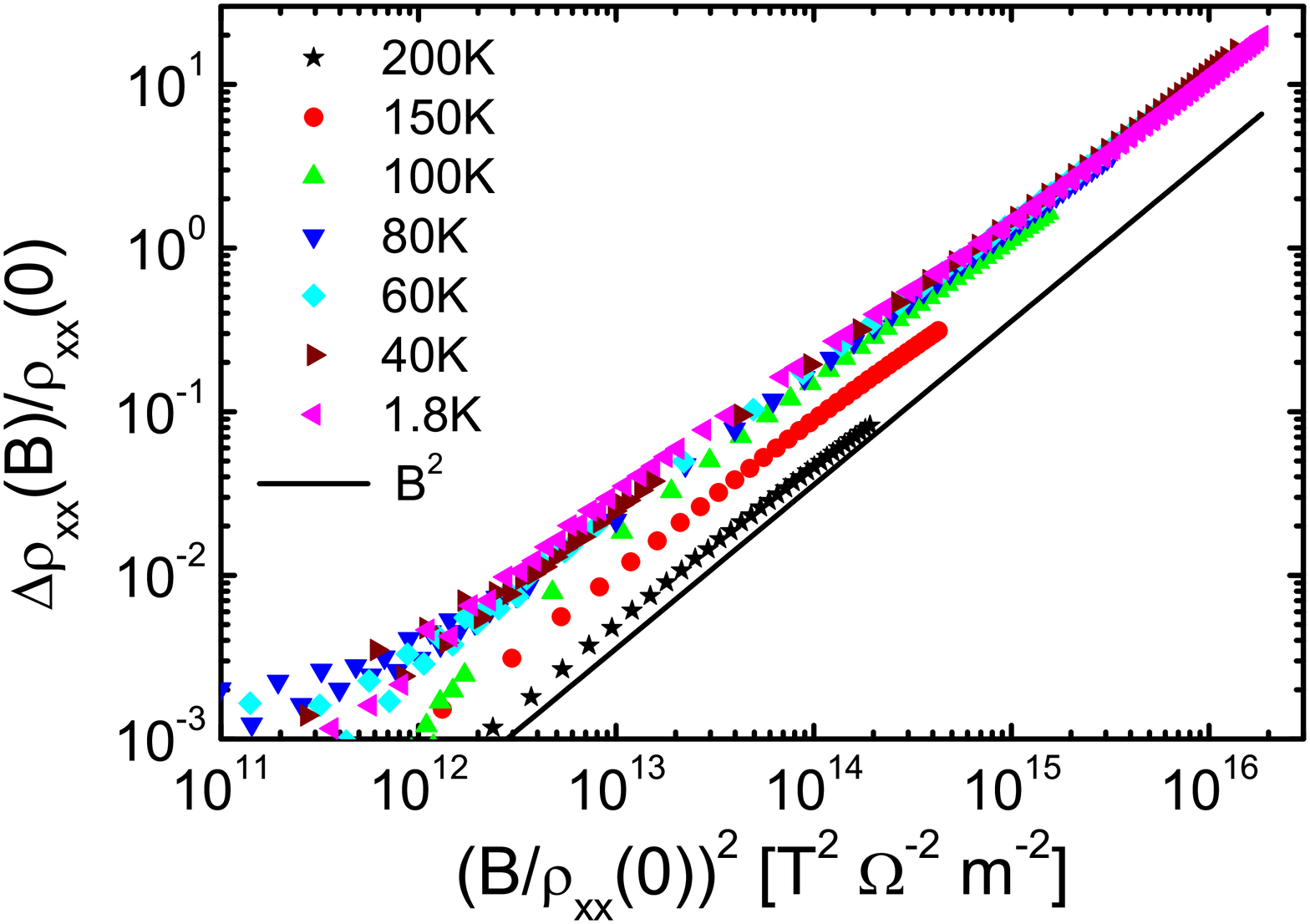}
   \caption{\label{Kohler} Magnetoresistance $\Delta\rho/\rho(B=0)$ plotted as a function of $B^{2}/\rho^2(B=0)$. Note the deviation from the Kohler's rule at high temperature.}    
\end{figure}

\begin{figure}
   \includegraphics[width=15 cm]{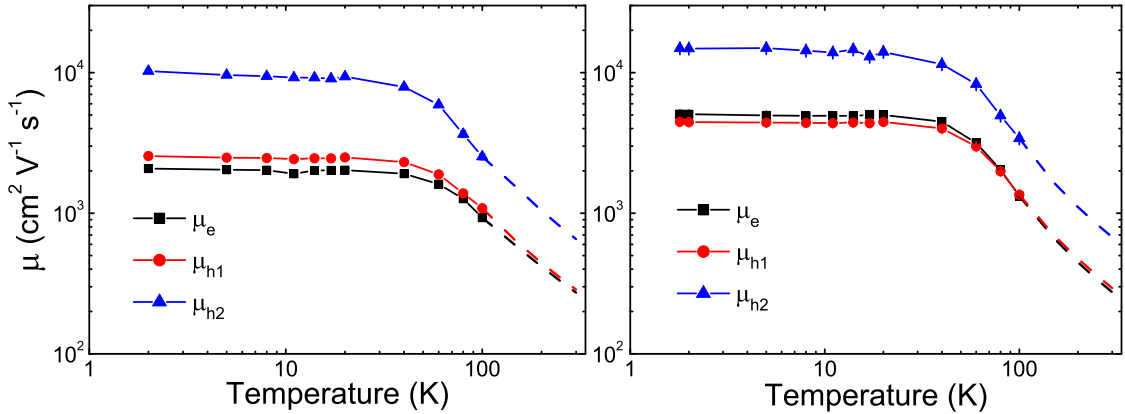}
   \caption{\label{mobility} (Color online) Carrier mobilities extracted from the analysis of the temperature and field dependence of the longitudinal and transverse conductivity for two BaNiS$_2$ samples with $RRR$ = 7.9 (left panel) and $RRR$ = 15.8 (right panel). The dashed lines above 100 K indicate the extrapolated behavior of the mobility by assuming a conventional $\mu_i \propto 1/T$ dependence expected for a metal with $\rho \propto T$.}    
\end{figure}

Fig. \ref{Mresistivity} displays the longitudinal and transverse components of the resistivity, $\rho_{xx}$ and $\rho_{yx}$ respectively, as a function of magnetic field. A sample with $RRR$ = 15.8 was chosen for this measurement. $\rho_{xx}$ exhibits a strong variation with $B$ which increases with decreasing temperature. The magnetoresistance $\Delta\rho_{xx}(B)/\rho_{xx}(0)= \left[\rho_{xx}(B)-\rho_{xx}(0)\right]/\rho_{xx}(0)$ does not exhibit any saturation up to 9 T and its large magnitude as high as 20 at 2 K is comparable to that of high mobility compounds, such as CuAgSe \cite{ish13}. A large magnetoresistance is typical of compensated semimetals like Bi \cite{kap28}, which is indeed the case of BaNiS$_2$, as indicated by the aforementioned band calculations and ARPES data. The field dependence of the transverse resistivity, $\rho_{yx}$, is linear at high temperature, as expected, but nonlinear at low temperature. In the high $RRR$ samples, the $\rho_{yx}$ curve displays a change of sign, a further signature of semimetallic properties. We further analyzed the magnetoresistivity data in order to determine the carrier density and carrier mobility. The plot of Fig. \ref{Kohler} put into evidence a deviation from Kohler's rule \cite{kohler}, which predicts a scaling of the $\Delta\rho_{xx}(B)/\rho_{xx}(0)$ curves as a function of the scaling variable $B/\rho_{xx}(0)$. This deviation is explained by the presence of at least two kinds of carriers with different mobilities. To investigate this point, we analyzed the temperature and field dependent data using the usual expressions for the longitudinal and tranverse magnetoconductivity:

\begin{figure}[!h]
   \includegraphics[width=10 cm]{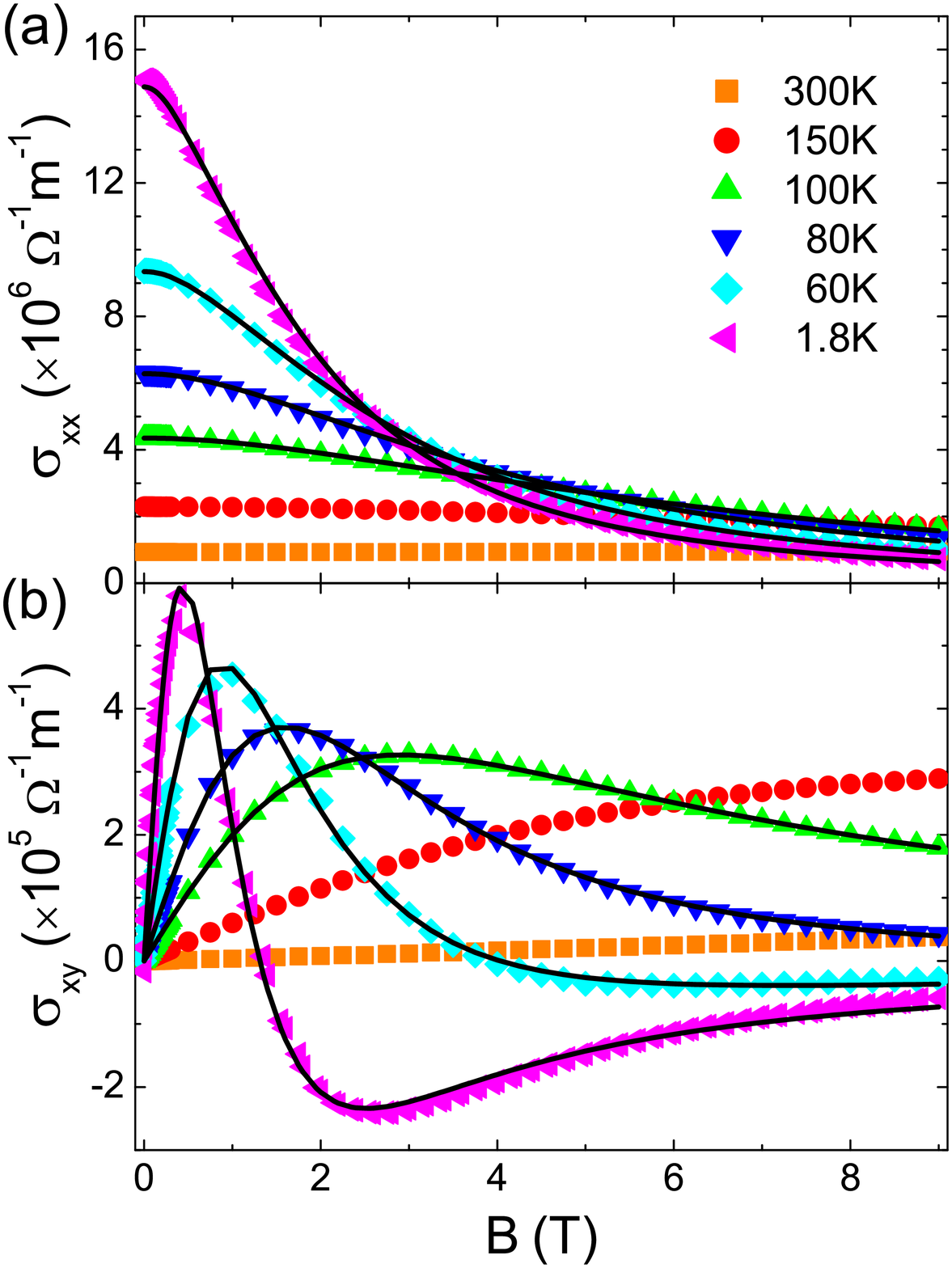}
   \caption{\label{Mconductivity} (Color online) Field-dependence of the components of the magnetoconductivity tensor, $\sigma_{xx}$ (a) and $\sigma_{xy}$ (b). Black solid lines are fits of the curves using Eqs. (\ref{xx},\ref{xy}) and a three-band model with two hole bands and one electron band.} 
\end{figure}

\begin{equation}
\sigma_{xx} = \sum_i \frac{|q_i|n_i\mu_i}{1+(\mu_iB)^2}
\label{xx}
\end{equation}

\begin{equation}
\sigma_{xy} = \sum_i \frac{q_in_i\mu_i^2B}{1+(\mu_iB)^2}
\label{xy}
\end{equation}
\\ 
where $q_i$, $n_i$ and $\mu_i$ are the charge, density and mobility of the carriers and $i$ is the band index. According to the above expressions, a significant field dependence of the conductivity appears if the term $\mu_iB$ is of the order of unity or larger. In this case, a simultaneous fit of the two components of the conductivity tensor by Eqs. (\ref{xx}) and (\ref{xy}) is possible, which allows a determination of $n_i$ and $\mu_i$ for each band. This is indeed our case, as the curves of Fig. \ref{Mconductivity} exhibit a strong field dependence up to 100 K. In this temperature range, we then tried to determine both $n_i$ and $\mu_i$ for each band $i$ by using Eqs. \ref{xx} and \ref{xy}. Whilst this analysis was not possible using a simple two-band model with one hole ($h$) and one electron ($e$) band, the inclusion of an additional hole-band yields a very good agreement with the data, as apparent from Fig. \ref{Mconductivity}. The fitting could not be extended to the data above 100 K, where the field-dependence of the magnetoconductivity is weak. The results of the data analysis are summarized in Table \ref{densities}. An indication of the reliability of this 3-band model is indicated by the fact that the inclusion of an extra electron - instead of hole - band was not successful. The carrier densities obtained from the above analysis are temperature and sample independent and fall in the typical $10^{19}-10^{20}$ cm$^{-3}$ range for semimetals, which account for the bad metallic properties of the compound. We find an excess of electrons of $\sim$1\% of the total number of carriers, which confirms the picture of compensated semimetal suggested by band calculations. In Fig. \ref{mobility}, the mobility of the three types of carriers are plotted as a function of temperature. While electrons and type-1 ($h_1$) holes exhibit relatively similar mobilities, the mobility of type-2 ($h_2$) holes is about 4-5 times larger and reaches the remarkable value of $\sim$ 15000 cm$^2$V$^{-1}$s$^{-1}$ in the cleanest crystals at low temperature. These values are comparable to the values $\sim$ 20000 cm$^2$V$^{-1}$s$^{-1}$ reported in undoped CuAgSe \cite{ish13} and only one order of magnitude less than in pure Si and GaAs semiconductors \cite{Rode75}, in spite of the much larger carrier densities in BaNiS$_2$.    

\begin{table}[!h]
\begin{center}
\begin{tabular}{@{}l| c c@{}}
\hline
\hline
$RRR$&7.9	&	15.8 \\
\hline
$n_e$ ($\times 10^{19}$ cm$^{-3}$)	&8.20(8) 	&8.75(9)\\
$n_{h1}$ ($\times 10^{19}$ cm$^{-3}$)	&7.07(14)	&7.4(4) \\
$n_{h2}$ ($\times 10^{19}$ cm$^{-3}$) &0.92(16) 	&1.17(12) \\
\hline
\hline
\end{tabular}
\end{center}
\caption{Carrier densities of two BaNiS$_2$ single crystals obtained from the analysis of the temperature- and field-dependent longitudinal and transverse conductivities by using eqs. (3,4), as explained in the text. The reported value are mean values of the carrier densities obtained at different temperatures. Number in parentheses indicate standard deviations.}
\label{densities}
\end{table}

\section{Discussion}

\begin{figure}
   \includegraphics[width=10 cm]{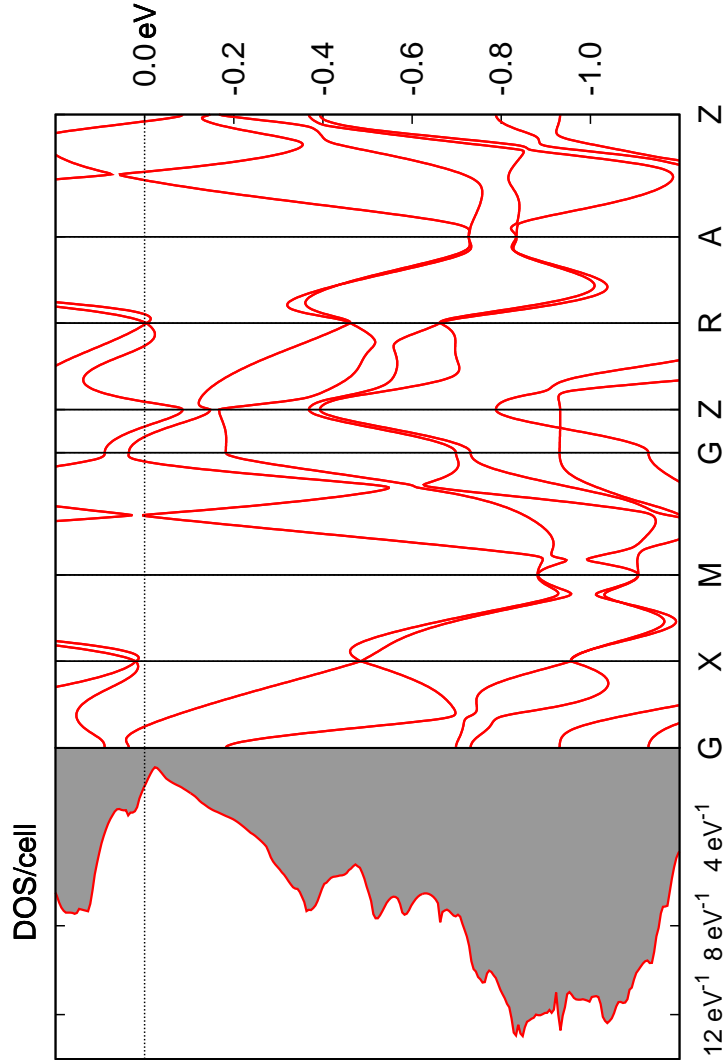}
   \caption{\label{bands} (Color online) Calculated electronic bands and total density of states. The dotted line indicates the Fermi level, $\epsilon_F$.}    
\end{figure}

We should first show that the three-band picture that emerges from the above analysis of the magnetoresistance data is consistent with the DFT band structure calculations reported in Figure \ref{bands} in the vicinity of the Fermi level, $\epsilon_F$. The bands crossing at $\epsilon_F$ and the Fermi sheets are as follows: 

\begin{enumerate}

\item Two linearly dispersive bands cross $\epsilon_F$ almost exactly at mid-distance along the $\Gamma M$ direction. This linear dispersion is in agreement with the observation of highly mobile $h_2$ holes. Due to their weak dispersion along $k_z$, these bands form a conic Fermi surface similar to the Dirac cone in graphene. In the present case, contrary to the case of graphene, the shape of the cone basis in the $k_x$-$k_y$ plane is not circular because of the different orbital character of the two bands that are mainly formed by $d_{3z^2-r^2}$ and $d_{x^2-y^2}$ states. In the light of the rich physics of the Dirac cones, it would be interesting to study the mobility of such $h_2$ holes as a function of doping.

\item The calculations predict a valence band and a conduction band at $\Gamma$ and Z respectively. The two bands, split by spin-orbit coupling effects, form the two hole and electron pockets visible along the $\Gamma$-Z direction, in agreement with the scenario of a second ($h_1$) hole band and of an electron band suggested by the previous analysis of magnetoresistance data.

\item In agreement with a very recent ARPES experiment \cite{ARPES}, two Rashba-split conduction bands are also visible at $R$. These bands touch the Fermi level and the volume of the resulting electron pockets is negligible as compared to that of the other pockets. Therefore, these bands should not affect the magnetotransport properties or marginally affect $\mu_e$ and $n_e$. 
\end{enumerate}

Second, we should discuss the anomalous properties that remain unexplained: (i) the levelling-off of the resistivity at values smaller than expected from the Mott-Ioffe-Regel limit; (ii) the crossover of resistivity regime at $T^{\ast}$ and (iii) the linear temperature-dependence of the susceptibility above 100 K.

Typically, feature (i) is the signature of a resistivity saturation in the Mott-Ioffe-Regel limit described by the condition that the electron mean free path becomes comparable to the Fermi wave length. In 2D, this limit is reached for $\rho_{sat}=2\pi\frac{e^2}{\hbar}c \approx$ 2.2 m$\Omega$cm. In our case, $\rho_{ab}$(400 K) is about one order of magnitude smaller than this value, so we consider the alternative explanation of a thermal activation of the carriers, which is significant in semimetals. Due to the low carrier density, $n$, which is apparent from the calculated density of states in Figure \ref{bands}, this activation leads to a large relative increase of $n$, detected as a reduction of the resistivity coefficient at high temperatures, as observed in graphite at $\sim$80 K \cite{His92}. In BaNiS$_2$, this reduction occurs at higher temperatures owing to the larger carrier density, $n \sim 10^{20}$ cm$^{-3}$, as compared to that of graphite, where $n \sim 10^{19}$ cm$^{-3}$ or less. This scenario is supported by the observation of a broad maximum of the thermoelectric power of BaNiS$_2$ at 175 K \cite{tak94}. The ARPES band structure showing that the bottom of the Rashba-split bands barely touch the Fermi level further suggests that these are the bands populated by thermal activation. Magnetotransport measurements in high magnetic fields above 100 K may confirm the validity of this scenario.\\

As to the anomaly of the resistivity at $T^{\ast}$, we first consider a scenario of \textit{umklapp} processes in combination with a singular DOS. It was previously proposed \cite{buh13} that, under these conditions, the electronic scattering rate acquires a non-analytic contribution leading to non-Fermi liquid transport properties down to very low temperatures. Specifically, the exponent of the temperature dependence of the resistivity can be significantly less than two \cite{sch10}. In the present case, the minimum of the Rashba-splitted bands at $R$ indeed forms a van Hove singularity near $\epsilon_F$. However, it is difficult to reconcile the large drop of the resistivity by a factor of two between $T^\ast$ and 40 mK with the modest number of minority carriers associated with this singularity. As a second possibility, we recall that the linear behavior of the resistivity is characteristic of correlated metals, such as heavy fermions in the vicinity of an AF order \cite{ste01} or superconducting cuprates \cite{coo09,hus11,and04}. However, in these systems, the linear dependence extends to a much wider temperature range, whilst in the present case this dependence is rather similar to a crossover between conventional Bloch-Gr\"uneisen regime above and low-temperature regime below $T^{\ast}$. Phenomenologically, this could be consistent with a crossover from a coherent Fermi-liquid regime to an incoherent regime controlled by a spin-freezing dynamics in the presence of a strong intra-atomic Hund coupling\cite{wer08}, which is expected in the present multiorbital case. According to this scenario, the loss of quasiparticle coherence at sufficiently high temperature leads to crossover of the transport regime \cite{geo13} in qualitative agreement with the present data.\\

As to anomaly (iii), it is recalled that a similar anomaly has been reported on Fe-based pnictides \cite{sko11}. For these systems, two scenarios have been proposed. The first scenario invokes the effect of the temperature-dependent chemical potential, $\mu$, near a sharp peak of the DOS \cite{Sal10}. As apparent from the usual expression for the Pauli susceptibility:

\begin{equation}
\label{Pauli}
\chi_{Pauli}=-\mu_{B}^{2}\int_0^\infty D(\epsilon)\frac{\partial f(\epsilon,\mu,T)}{\partial \epsilon}d\epsilon
\end{equation}
\\
where $f(\epsilon,\mu,T)$ is the Fermi-Dirac distribution and $\mu_B$ is the Bohr magneton, a linear dependence of $\chi$ appears if the DOS strongly varies with energy near $\epsilon_F \approx \mu$, as in the case of semimetals. This simple picture has been proposed for pnictides on the basis of dynamic mean field theory (DMFT) calculations \cite{sko11,sko12}. At first sight, a similar picture may be applicable to the present case as well, for the calculated band structure of BaNiS$_2$ also exhibits a pronounced dip in the DOS at $\sim 0.1$ eV below $\epsilon_F$. Indeed, the calculated $\chi(T)$ reproduce qualitatively the observed behavior of $\chi(T)$, including the existence of a tail at low temperature and the linear term at high temperatures. However, the experimental slope $d\chi(T)/dT$ is about one order of magnitude larger than predicted using Eq. \ref{Pauli} and the calculated DOS. One may consider the possibility that the real DOS is much more singular than the calculated one due to renormalization effects. However, this possibility contrasts the evidence of modest renormalization effects from the present specific heat results and from ARPES measurements \cite{ARPES}.

A second scenario invokes the existence of short-range AF fluctuations \cite{kli10,wan09}, in agreement with previous theoretical studies of frustrated Heisenberg models with comparable nearest-neighbor and next-nearest-neighbor exchange coupling constants, $J_1$ and $J_2$ \cite{mil91}. This scenario could be applicable to the present case since a $J_1$-$J_2$ model is consistent with the observation of collinear AF order in BaCoS$_{2}$ \cite{man97} and with the fact that the next-nearest-neighbor Ni-Ni distance in the square lattice of BaNiS$_{2}$ is only $\sqrt{2}$ times the nearest-neighbor distance. Thus, it is envisaged that residual short-range fluctuations may be present in BaNiS$_{2}$ as well. Suitable studies of spin dynamics would be required to probe directly these fluctuations. 

\section{Conclusion}
In conclusion, the present study on high-quality BaNiS$_{2}$ single crystals suggests a consistent picture of compensated semimetal with two hole bands and one electron band, in agreement with ARPES data and DFT calculations. Notable is the high mobility of the minority holes attributed to a Dirac-like Fermi pocket, similar to the case of graphene. The transport properties are markedly two-dimensional and dominated by a conventional electron-phonon mechanism. While the low-temperature electronic specific heat is consistent with the calculated density of states, the magnetic susceptibility is strongly enhanced and exhibits a pronounced linear term at high temperature. This anomaly could be explained either by a strong variation of the density of state at $\epsilon_F$, not predicted by DFT calculations, or by the presence of short-range AF fluctuations reminiscent of the AF order observed in BaCoS$_{2}$. A further signature of unconventional behavior is a striking crossover of the resistivity from a conventional Bloch-Gr\"uneisen behavior to a linear behavior below $T^{\ast} \sim 6$ K. We envisage that this crossover may reflect the loss of quasi-particle coherence caused by a spin-freezing dynamics in presence of a large Hund coupling. Further studies are required to verify the validity of the above scenario according to which both intra- and inter-atomic magnetic interactions strongly alter the stability of a conventional Fermi-liquid ground state.

\begin{acknowledgments}
This work was partly supported by the University Pierre and Marie Curie under the \textit{Programme \'emergence} funding program and by the IDRIS/GENCI computational resources under the project No. 096493. We are grateful to B. L\'eridon and R. Lobo for giving us access to the VSM-SQUID apparatus of their laboratory supported by the \textit{R\'egion \^Ile-de-France}. We are thankful to J. Biscaras for fruitful discussions.   
\end{acknowledgments}

% Create the reference section using BibTeX:
\bibliography{biblio}
\end{document}